\documentstyle[epsfig]{aipproc}
\newcommand{\beq}{\begin{equation}}
\newcommand{\eeq}{\end{equation}}
\newcommand{\bea}{\begin{eqnarray}}
\newcommand{\eea}{\end{eqnarray}}
\newcommand{\apj}{{\it Astrophys. J.} }
\begin{document}
\title{Self-Similar Hot Accretion Flow onto a Rotating Neutron Star: 
Structure and Stability}

\author{Mikhail V. Medvedev$^{1}$ and Ramesh Narayan$^2$ }
\address{
$^1$CITA, University of Toronto, Toronto, Ontario, M5S 3H8, Canada\\
$^2$Harvard-Smithsonian Center for Astrophysics, 60 Garden Street,
Cambridge, MA 02138}

\maketitle

\begin{abstract}
We present analytical and numerical solutions which describe a hot,
viscous, two-temperature accretion flow onto a rotating neutron star or any
other rotating compact star with a surface.  We assume Coulomb coupling between
the protons and electrons, and free-free cooling from the electrons.
Outside a thin boundary layer, where the accretion flow meets the
star, we show that there is an extended settling region which is
well-described by two self-similar solutions: (i) a two-temperature
solution which is valid in an inner zone $r\le10^{2.5}$ ($r$ is in
Schwarzchild units), and (ii) a one-temperature solution at larger
radii.  In both zones, $\rho\propto r^{-2},\ \Omega\propto r^{-3/2},\
v\propto r^0,\ T_p\propto r^{-1}$; in the two-temperature zone,
$T_e\propto r^{-1/2}$.  The luminosity of the settling zone arises
from the rotational energy of the star as the star is braked by
viscosity. Hence the luminosity and the flow parameters (density,
temperature, angular velocity) are independent of $\dot M$. The settling
solution described here is not advection-dominated, and is thus
different from the self-similar ADAF found around black holes.  When
the spin of the star is small enough, however, the present solution
transforms smoothly to a (settling) ADAF. 

We carried out a stability analysis of the settling flow. The flow is 
convectively and viscously stable and is unlikely to have strong winds 
or outflows.  Unlike another cooling-dominated system --- the SLE disk, ---
the settling flow is thermally stable provided that thermal conduction 
is taken into account. This strong saturated-like thermoconduction does
not change the structure of the flow.
\end{abstract}

\section*{The Settling Flow}

At small mass accretion rates, $\lesssim10^{-2}$ of the Eddington
rate, black holes (BHs) accrete via an ADAF --- a hot, two-temperature, 
radiatively inefficient, geometrically thick, advection-dominated accretion 
flow \cite{NY94,NY95}. In contrast, accretion onto compact stars, e.g.,
a neutron star (NS) may occur via either an ADAF of a settling solution
\cite{MN01}. The latter corresponds to strongly rotating stars only.
In the settling flow the rotational energy is extracted from the star 
via viscous torques in the boundary layer where the flow meets the star
surface. The extracted energy heats the flow and ultimately escapes from
the flow as free-free radiation. In addition, viscosity extracts angular
momentum from the star as well. In the stationary flow, this angular
momentum must be transported though the flow to the outermost 
radii, where it goes into the ambient medium. It is this huge angular
momentum flux $\dot J$ which modifies the entire structure of the accretion
flow and makes it drastically different from an ADAF. Note that no angular
momentum may be extracted from a BH horizon by viscosity. Thus a viscous
settling flow may exist in NS systems and not in BH systems.

\begin{figure}[b!] 
\centerline{\epsfig{file=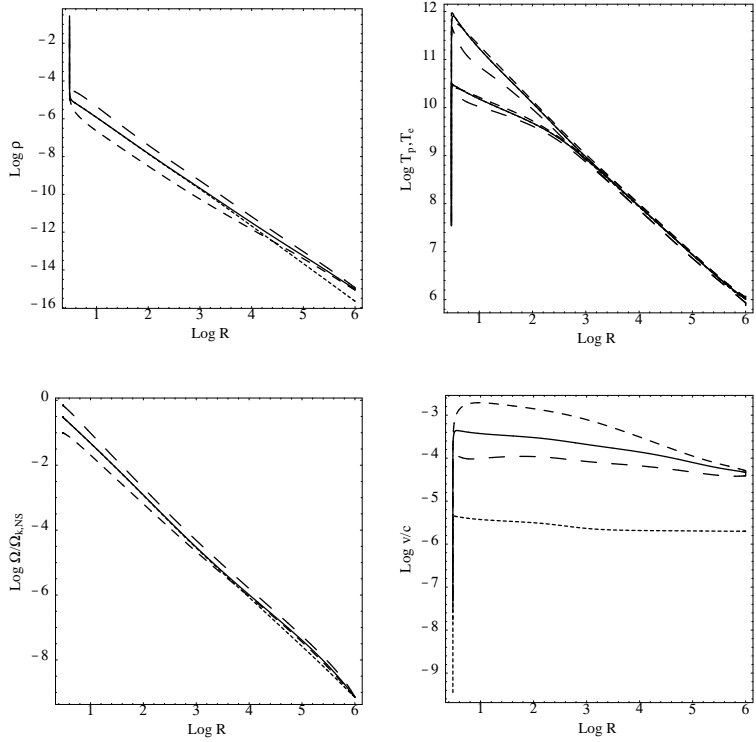,width=4.1in}}
\vspace{10pt}
\caption{Profiles of density $\rho$ (g~cm$^{-3}$),
proton temperature $T_p$ ($^\circ$K), electron temperature $T_e$
($^\circ$K), angular velocity $\Omega$ (in units of the Keplerian
angular velocity at the NS radius $R_{NS}$), and radial velocity $v$ (in
units of $c$) for accretion flows with $\alpha=0.1$ and $(\dot m,s$) =
(0.01,0.3) -- solid line, (0.0001, 0.3) -- short-dashed line,
(0.01,0.1) -- medium-dashed line, (0.01,0.7) -- long-dashed line.}
\label{fig1}
\end{figure}
The structure of the steady, rotating, axisymmetric, quasi-spherical,
two-temperature settling flow has been found analytically and confirmed 
numerically \cite{MN01}. We use the height-integrated form of the viscous 
hydrodynamic equations with the Shakura-Sunyaev-type viscosity parametrized 
by dimensionless $\alpha$. We assume viscous heating of protons, Bremsstrahlung 
cooling of electrons and Coulomb energy transfer from the protons to the 
electrons; we neglect Comptonization but include thermal conduction in 
the form discussed in the next section. In the inner zone $r<10^{2.5}$
($r$ is in Schwarzchild units, $R_S=2GM/c^2$), the flow is two-temperature
with the density, proton and electron temperatures, angular and
radial velocities scalings as 
\beq
\rho=\rho_0\,r^{-2},\quad T_p=T_{p0}\,r^{-1},\quad 
T_e=T_{e0}\,r^{-1/2}, \quad \Omega=\Omega_0\,r^{-3/2},\quad v=v_0\,r^0, 
\label{sss}
\eeq
where $\rho_0,\ T_{p0},\ T_{e0}, \Omega_0, v_0$ are functions
of $M,\ \alpha$ and the star spin $s={\Omega_*/\Omega_K(R_*)}$, and
$\Omega_K(R)=(GM/R^3)^{1/2}$ is the Keplerian angular velocity.
In the outer zone $r>10^{2.5}$, we have $T_e=T_p\propto r^{-1}$ and the
same other scalings. This self-similar solution is valid for the
part of the flow below the radius $r_s$ related to the mass accretion
rate $\dot m$ (in Eddington units, $\dot M_{\rm Edd}=1.4\times10^{18}m
\textrm{ g/s}$, and here $m=M/M_{\sun}$):
\beq
\dot m<2.2\times10^{-3}\alpha_{0.1}^2s_{0.3}^2r_{s,3}^{-1/2},
\label{out-constr}
\eeq 
where $r_{s,3}=r_s/10^3$,\ $\alpha_{0.1}=\alpha/0.1$, etc..
The numerical solution of the hydrodynamic equations with appropriate 
inner and outer boundary conditions is represented in Figure \ref{fig1}.
It is in excellent agreement with the self-similar solition (\ref{sss}).
\begin{figure}[b!] 
\centerline{\epsfig{file=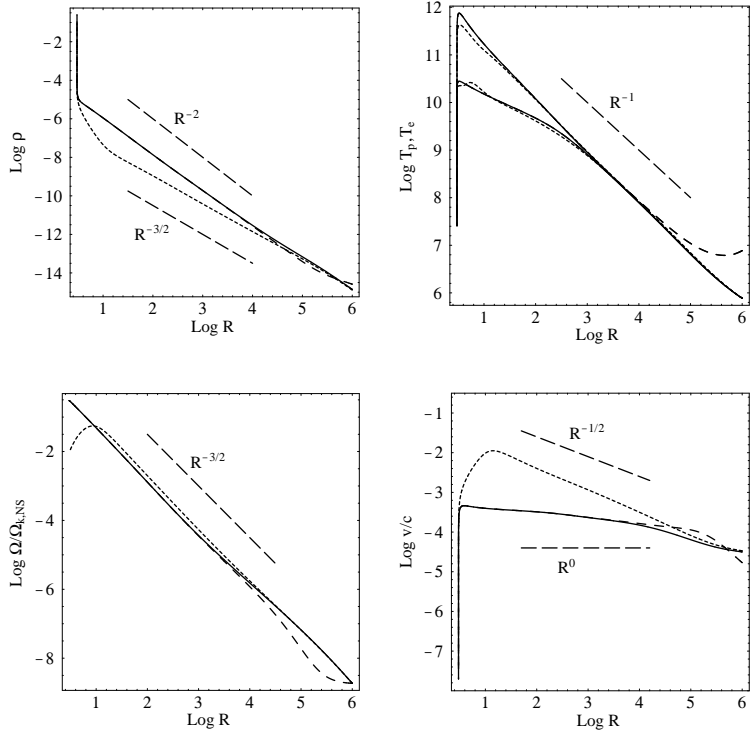,width=4.1in}}
\vspace{10pt}
\caption{Same as in Fig.\ \ref{fig1} for $\gamma=4/3$ and 
$s=0.3$ (solid curve) and $s=0.01$ (dotted curve). The self-similar 
slopes for an ADAF flow and a settling flow are shown for comparison.  
The long-dashed curves represent the same solution as the solid curve, 
but with ten times higher temperature at $R_{out}$.}
\label{fig2} 
\end{figure}

There is a remarkable property of the settling flow: all quantities,
e.g., $\rho,\ T_p$, etc., except $v$, are {\em independent} of the
mass accretion rate $\dot M$. This happens because the gravitational
energy of the accreting gas is much smaller than the energy extracted
from the rotating star. It is this energy which dominates the flow
luminosity. We should also remark that the settling flow is more similar 
to a steady, radiative cooling-dominated ``atmosphere'' rather than to a 
rapidly infalling flow: the radial infall velocity is constant and is much less 
then the free-fall velocity $v/v_{\rm ff}\propto r^{1/2}\to 0$ as $r\to0$. 
The structure of the settling is very sensitive to the rotation rate 
of the central star. As the angular velocity decreases below few
percents of the Keplerian value, the settling flow smoothly transforms
into a conventional ADAF solution, as represented in Figure \ref{fig2}. 

It was shown that the settling flow is (i) convectively stable
and (ii) may not have strong winds and outflows (the Bernoulli 
number is negative) if the adiabatic index satisfies
\beq 
\gamma>\frac{3\left(1-s^2/2\right)}{\left(2-s^2/2\right)}\sim1.5 .
\label{b<0}
\eeq 
Other properties of the settling flow are discussed elsewhere
\cite{MN01,Mtexas00}.

\section*{Thermal Stability of the settling flow}

The settling flow is cooling-dominated. Thus it is similar to the
Shapiro-Lightman-Eardley solution \cite{SLE76} which is known to 
be thermally unstable \cite{P78}. Hence, our settling solution may
be unstable as well. To study the thermal instability in sheared circular 
accretion flows we use the shearing sheet approximation with the velocity 
given by
\beq
{\bf V}_0(x)=2A\,x\,\hat{y},
\eeq
where $2A=d{\bf V}_0/dx$ is the shear frequency, $x,\ y,\ z$ are the
radial, azimuthal, and vertical coordinates, and ``hat'' denotes
a unit vector. We assume that there is a Coriolis acceleration
due to ${\bf \Omega}=\Omega\,\hat{z}$. The vorticity and epicyclic
frequency are then $2B=2A+2\Omega,\ \kappa_{\rm epi}^2=4\Omega B$.
For a Keplerian-type flow, $\Omega\propto R^{-3/2}$, which is the case
for both the settling and SLE solutions considered below, one has
$2A=-(3/2)\Omega$ and $2B=\Omega/2$.
We assume that perturbations have only $s$-component, which
corresponds to axisymmetric perturbations. We ignore motions
in $z$ direction. We use hydrodynamic equations with thermoconductive flux.

The settling flow is hot (sub-virial), so that the mean-free-path (of both
electrons and protons) is larger than the size of the system. Hence the
conventional Spitzer theory fails. Without collisions but in
the presence of magnetic fields electrons stream freely along the field lines, 
therefore the parallel heat flux remains large. In contrast,
transverse heat flux is greatly reduced in a magnetic field because electrons
are tied to the field lines on the scale of the Larmor orbit and cannot move 
across the field lines too far. In a tangled field, however, electrons can 
jump from one field line to another and thus transfer heat across the field 
\cite{RR78}. The average thermal conductivity in tangled fields is
\beq
\kappa_B\simeq nk_Bv_Tl_B\,\vartheta 
\simeq 10^{-2}nk_Bv_TR\xi_{-1}\vartheta_{-1},
\label{kappa_B-am}
\eeq
where $l_B=\xi R$ is the correlation scale of the magnetic fields set by
flow turbulence and $\vartheta<1$ takes in to account that only a fraction
of all particles may pass though the magnetic barriers. The rest of them will 
remain  trapped in magnetic wells and, hence, will not transport energy to 
large distances $\gg l_B$. The thermoconductive flux is 
\beq
q_{\rm cond}=-\alpha_c\frac{\rho c_s^2}{\Omega_K}\frac{dc_s^2}{dx},
\label{q-cond}
\textrm{~~~where~~~}  
\alpha_c
\simeq\frac{R}{H}\,\xi\vartheta\, F(e,p)
\simeq10^{-2}\xi_{-1}\vartheta_{-1}\, F(e,p)
\label{alpha_c}
\eeq
is the ``alpha-prescription'' thermo-conductivity coefficient and 
$F(e,p)\ [1\le F(e,p)\lesssim15]$ takes into account that conduction may be 
dominated by the protons or the electrons, depending on the flow conditions.
Here we used that $v_T\simeq c_{se}$ and $H/R\sim c_s/v_{\rm ff}
\sim c_s/\Omega_KR$, where $H\sim R$ is the accretion disk scale height.

Now, from hydrodynamic equations for the perturbations, we obtain the 
dispersion relation
\bea
& &{}\omega\left[\frac{\omega}{\gamma-1}+\frac{i(n-1)}{\tau_{\rm cool}}
+\frac{ik^2R^2}{\tau_{\rm cond}}\right] 
\left(\omega^2-\kappa_{\rm epi}^2-k^2c_s^2\right)
-\omega\left[\omega+\frac{i(2B/A-1)}{\tau_{\rm cool}}\right]k^2c_s^2=0,
\label{disp} \nonumber\\
& &{}\textrm{where}\qquad
\tau_{\rm cool}=\frac{4}{(9\alpha s^2+2\alpha_c)}\,\Omega_K^{-1}
\simeq\frac{2}{\alpha_c}\,\Omega_K^{-1}, \qquad
\tau_{\rm cond}=\frac{3}{\alpha_c}\,\Omega_K^{-1}.
\eea
In the large-$k$ limit, (\ref{disp}) yields the growth rate of the 
thermal mode. The stability criterion ${\rm Im}\,\omega<0$ is 
(for $\alpha s^2 \lesssim \alpha_c$)
\beq
kR>\left[\frac{\tau_{\rm cond}}{\tau_{\rm cool}}\left(2-n-2\,\frac{B}{A}\right)
\right]^{1/2}
=\left[\frac{26\alpha_c}{9(9\alpha s^2+2\alpha_c)}\right]^{1/2}
\simeq\sqrt{\frac{13}{9}}\simeq1.2,
\label{stab}
\eeq
that is, thermal modes with $kR\ge2$ are stable.
Whether the mode $kR=1$ is stable or not cannot be reliably determined
from our local approach. A global stability analysis is necessary to
properly account for the effects of geometry and curvature on the 
eigenmode structure. Finally, we conclude that the settling flow
is very likely stable.

\end{document}